\documentclass[11pt]{article}
\usepackage{graphicx,verbatim,array,multicol,palatino,amssymb,amsfonts, amsmath, subfigure, epsfig,multirow,caption}
\usepackage{color}
\usepackage{psfrag}
\usepackage{chicago}
\usepackage{epstopdf}
\usepackage{pdflscape}

\def\lboxit#1{\vbox{\hrule\hbox{\vrule\kern6pt
      \vbox{\kern6pt#1\kern6pt}\kern6pt\vrule}\hrule}}

\def\thick#1{\hbox{\rlap{$#1$}\kern0.25pt\rlap{$#1$}\kern0.25pt$#1$}}

\def\bphi{{\bf \phi}}

\newtheorem{prop}{Proposition}

\begin{document}
\title{Relabelling Algorithms for Large Dataset Mixture Models}
\author{W. Zhu \: and
Y. Fan\footnote{School of Mathematics and Statistics, University of New South Wales, Sydney 2052 Australia. Communicating Author Y. Fan: Email \tt{Y.Fan@unsw.edu.au.}}
}


\maketitle

\begin{abstract}
 Mixture models are flexible tools in density estimation and classification problems. Bayesian estimation of such models typically relies on
 sampling from the posterior distribution using Markov chain Monte Carlo. Label switching arises because the posterior is invariant to
 permutations of the component parameters. Methods for dealing with label switching have been studied fairly extensively in the literature, with the most
 popular approaches being those based on loss functions. However, many of these algorithms turn out to be too slow in practice, and can be infeasible
 as the size and dimension of the data grow. In this article, we review earlier solutions which can scale up well for large data sets, and compare their
 performances on simulated and real datasets. In addition,
 we propose a new, and computationally efficient algorithm based on a loss function interpretation,  and show that it can scale up well in larger problems. 
 We conclude with some discussions and recommendations of all the methods studied.

\end{abstract}

\section{Introduction}

Mixture models have been used extensively in statistics, in areas such as nonparametric density estimation (\shortciteNP{norets2010approximation}) and
model based clustering (\shortciteNP{banfield1993model}, \shortciteNP{mclachlan1988mixture}). These models provide a flexible way of modelling heterogeneous data. Here we are concerned with
finite mixture distributions of $K$ components with density given by
\begin{equation}\label{equ1.1}
 p(x_i|  {\bphi})=\sum_{k=1}^{K}w_{k}f(x_i \mid {\bf \theta}_{k})
\end{equation}
for some data $x_i \in {\mathcal R}^d, d\geq 1, i=1,\ldots, n$. $f(x_i | {\bf \theta}_k)$ is the $k$th component density of the mixture, with
parameters ${\bf \theta}_k$. For instance, $f(x_i | {\bf \theta}_k)$  can be an univariate or a multivariate Normal distribution where the parameter vector ${\bf \theta}_k$ represents the mean and variance/covariance of the Normal distribution. Finally $w_k$ is the weight of the $k$th component density, such that
$\sum_{k=1}^K w_k=1$. We will denote the entire $q$-dimensional set of parameters as $\phi = ((w_1, \theta_1), \ldots, (w_K, \theta_K))$. Comprehensive reviews of finite mixture models can be found in \shortciteN{titterington1985statistical}, \shortciteN{mclachlan2004finite}, \shortciteN{marinmr05}, \shortciteN{schnatter06} .
\\

Bayesian analyses of finite mixture models typically involve the use of Markov chain Monte Carlo (MCMC) sampling from the posterior distribution, where label switching becomes
an issue which requires resolving. This occurs as a result of the invariance of the Equation (\ref{equ1.1}) with respect to the reordering of the components such that
\[
\sum_{k=1}^{K}w_{k}f(x_i \mid {\bf \theta}_{k}) = \sum_{k=1}^{K}w_{\nu_k}f(x_i \mid {\bf \theta}_{\nu_k})
\]
where $\{\nu_1,\ldots, \nu_K\}$ is some arbitrary permutation of $\{1,\dots, K\}$. If the priors of the parameters are the same or exchangeable, the posterior distribution will be invariant under the permutation.
One can visualise the occurrence of label switching within an MCMC sampler, when for instance the parameters of the first component moves  to the
modal region of the second component as the Markov chain explores the state space, and vice versa. While the posterior density remains invariant to the labelling, the correct ordering of the labels should have swapped the two sets of parameters.
\\

Many methods have been developed to resolve the issue of identifiability in Bayesian inference. \shortciteN{jasraHS05}  provides a detailed and insightful review of developments on this topic up to around 2005. The simplest method is to impose an artificial identifiability constraint. For instance, \shortciteN{richardson1997bayesian} suggest ordering the location parameters of a univariate Normal mixture model, such that $\mu_1  < \ldots <\mu_K$, where $\mu_k$ corresponds to the mean parameter of the $k$th component. Imposing such identifiability constraints  can also be seen as a modification of the prior distribution. The method is simple regarding computational complexity. and efficient, it can also be implemented online within the MCMC sampler.
However, it was demonstrated in \shortciteN{jasraHS05} and \shortciteN{celeuxhr00} that the method can fail to fully resolve the issue of identifiability in some cases. Additionally, in higher dimensional problems, it becomes difficult to know how to set the identifiability constraint,
see \shortciteN{schnatter11}  for an example in the case of  multivariate Normal mixtures.
\\

Another class of relabelling algorithms, perhaps the best known algorithms in the literature to date,  is based on decision theoretic arguments. Samples from MCMC output are post-processed according to some loss function criterion, see \shortciteN{stephens97}, \shortciteN{stephens00}, \shortciteN{celeux98}, \shortciteN{celeuxhr00}, \shortciteN{hurnjr03} and references therein. These methods can work well, and are considered to be theoretically better justified by   \shortciteN{jasraHS05}. However, they are computationally intensive, thus for large datasets or high dimensions, they become impractical to use.
\\

Finally, a different approach, based on probabilistic relabelling, can be found in the works of  \shortciteN{sperrinjw09} and \shortciteN{jasra05}, which
involve the calculation of the likelihood of the permutations $\{\nu_1,\ldots, \nu_K\}$.   \shortciteN{sperrinjw09} gives an EM-type algorithm for its estimation. \shortciteN{puoloamakiK09} develop a relabelling approach which requires the introduction of a discrete latent variable in the original probabilistic model. More recently, \shortciteN{yaol09} propose an algorithm based on the mode of the posterior and an ascent algorithm for each iterate of the MCMC sample, \shortciteN{yaol12} propose a method  which minimizes the class probabilities to a fixed reference label, \shortciteN{yao12} proposes to assign the probabilities for each possible labels by fitting a mixture
model to the permutation symmetric posterior. While many of these algorithms were demonstrated to work well, they do not scale well for large data or high dimensions.\\

Many modern applications of mixture models involve increasingly large datasets and in higher dimensions, such as those found in genetic studies, and medical image analyses. Here, we are focused on efficient relabelling algorithms which can scale well to large data and high dimensional problems.  We first review
existing algorithms in this category in Section \ref{sec:review}, and then in Section \ref{sec:minvar}, introduce a new algorithm which is interpretable under the squared loss function. We extensively compare these algorithms in Section \ref{sec:example}, and conclude with some discussions and recommendations in Section \ref{sec:discussion}.

\section{Review of existing relabelling algorithms} \label{sec:review}
In this section, we focus our review on relabelling methods which can handle high dimensional problems, and those which will scale up well for large data sets. In addition, readers are referred to the excellent review of \shortciteN{jasraHS05} for a more general review for developments prior to 2005, here we will focus more closely on scalable algorithms to higher dimesions. \\

We broadly separate the class of relabelling algorithm into two categories, one which works on the full set of $q$-dimensional parameters $\phi$, and refer to these as full parameter space relabelling. A second category works on the allocation parameters only. We shall refer to these as the allocation space relabelling algorithms.

\subsection{Full parameter space relabelling algorithms}\label{sec:fullpar}


\subsubsection{Celeux et al (1998, 2000)}
\shortciteN{celeux98} and \shortciteN{celeuxhr00} provide a simple algorithm for relabelling. Here, a reference modal region is selected using the initial iterations of the MCMC sampler, and subsequent points are then permuted with respect to the reference points, according to a $k$-means type algorithm. \\

Let $\phi^{j} = ((w^j_1, \theta_1^j), \ldots, (w^j_K, \theta_K^j)) $ denote the vector of parameter estimates at the $j^{th}$ iteration of the MCMC sampler.
Initialise with the first $m$ sample outputs, where $m$ is sufficiently
large to ensure that the initial estimates are a reasonable approximation to the posterior means, but not so large that label switching has already occurred. \shortciteN{celeuxhr00} suggests that $m=100$ is typically sufficient. Then define component specific location and scale measures
$$\bar{\phi}_i = \frac{1}{m}\sum_{j=1}^m \phi_i^j$$
and
$$s_i=\frac{1}{m}\sum_{j=1}^m (\phi_i^j-\bar{\phi}_i)^2$$
for $i=1,\ldots,q$. Then treating this as the initial ordering, $K! -1$ other location and scale labels are produced from this set, thus we denote the
entire initial set of permutations of location and scale values by $\{\bar{\phi}_{\nu_{k}}^{[0]},  s_{\nu_{k}}^{[0]}\}$, where $\nu_k$ denotes the set of all possible permutations.\\

Subsequent iterations of the relabelling algorithm then proceeds by allocating the permutation $\nu_{k^*}$ to the $m+r^{th}$ MCMC output vector $\phi^{m+r}$
which minimises the scaled Euclidean distance of all components $i=1,\ldots, q$, namely we find the permutation  $\nu_{k^*}$
$$\nu_{k^*}=\underset{\nu_{k}}{\mbox{argmin}} \sum_{i=1}^q \frac{\phi_i^{m+r} - \bar{\phi}_{\nu_{k}, i}^{[r-1]}}{s_{\nu_{k},i}^{[r-1]}},$$
where $\bar{\phi}_{\nu_{k}, i}^{[r-1]}$ and $s_{\nu_{k},i}^{[r-1]}$ are respectively the $i^{th}$ coordinate of the current estimate of the location and scale vector with respect to the permutation $\nu_k$. Finally, the location and scale vectors are updated with the new $r^{th}$ sample.\\

This algorithm works by minimising the scaled Euclidean distance to the cluster centers, assuming the initial centers provided a good estimate.
In practice, the use of component variance for scaling, leads to those components with very small variances dominating the others, hence leading to
inaccurate relabelling in these types of problems, as demonstrated in our simulation studies in later sections.


\subsubsection{Fr{\" u}wirth-Schnatter (2011)}
\shortciteN{schnatter06} and \shortciteN{schnatter11} propose to apply the standard $k$-means algorithm with $K$ clusters to all the MCMC sample output, with the posterior mode estimator $\phi_1^*,\ldots, \phi_K^*$ serving as starting value for the cluster means. They suggest that each element of the parameter vector should be standardised.\\

If the simulation clusters are well separated, then the classification sequence given by the classification index is a permutation, that is, the $k$-means algorithm allocates each component parameter vectors to exactly $K$ clusters.  However, this is not always the case, and the algorithm can often allocate multiple components to the same cluster. \shortciteN{schnatter11} suggest that a simple check by ordering of the sequence of classification index, and if this does not equal $\{1,\ldots,K\}$ then the sample is simply excluded. \\


The algorithm is very simple and efficient, it is easy to understand as it uses the well known $k$-means clustering algorithm. However it
 can become inefficient when cluster components are very close to each other, leading to allocation of multiple components into the same cluster.
 Since such samples are then excluded for analyses, this can result in high proportion of waste of  MCMC samples, which can themselves be expensive to calculate in high dimensional problems.

\subsection{Marin et al (2005)}

\shortciteN{marinmr05} provides a simple algorithm for the reordering of MCMC output of size $M$, they first find the posterior mode $\phi^{*}$, then for
each sample, compute
$$\nu_{k^*}=\underset{\nu_{k}}{\mbox{argmin}} <\phi_{\nu_k}, \phi^*>_q,$$
where $<>_q$ is the canonical scalar product of ${\mathcal R}^q$. \\

Thus each MCMC output is reordered with respect to the approximate posterior MAP estimator. Several authors, e.g. \shortciteN{jasraHS05}  and \shortciteN{papastami10} comment on the simplicity of the method, but note that it can fail when there's multimodality in the parameters.

\subsection{Allocation space relabelling algorithms}
In the allocation sampler  (see \shortciteN{nobilef07}),  a latent variable ${\bf z}$, is introduced for each observation, which indicates the component membership. This approach
is often used when clustering observations into different subsets is the aim.  Relabelling based on allocation variable alone has the advantage that its computational cost is invariant to increases in the dimensionality of the parameter space.  \\

The allocation sampler is obtained by augmenting Equation \eqref{equ1.1} with the auxiliary variable ${\bf z} = (z_1,\ldots, z_n)$, such that
$$p(z_i = k) =w_k, \mbox{ for } k =1,\ldots, K,$$
and 
$$p(x_i|  {\bphi}, z_i)= f(x_i \mid {\bf \theta}_{k}, z_i), $$
so that 
\begin{equation}\label{equ1.2}
 p(x_i|  {\bphi})=\sum_{k=1}^{K}w_{k}f(x_i \mid {\bf \theta}_{k}, z_i).
\end{equation}

Note that when the allocation sampler is not used,  the algorithms in this section can be used by computing a plug in estimate of the allocation for each MCMC
iteration $j$, 
\begin{equation}\label{allocate}
\hat{z}^j_i=  \underset{k}{\mbox{argmax }}   w_k f(x_i|\phi^j, z_i=k)/ p(x_i | \phi^j),
\end{equation}
similar approaches can be found in for example, \shortciteN{stephens00}.

\subsubsection{Cron and West (2011) }
\shortciteN{cronw11} provides a relabelling algorithm based entirely on the latent variables.
Define $\hat{z}$ to be the vector with $n$ elements $\hat{z}_i$, which either arises naturally via the allocation sampler as in Equation \eqref{equ1.2}, or it  can be determined according to Equation \eqref{allocation}. So $\hat{z}$ assigns each data 
observation to its modal component under the current set of classification probabilities. Define $\hat{z}^R$ as the classification vector with elements $\hat{z}_i^R$ at some reference point, ideally taken as the posterior mode. They suggest a Bayesian EM algorithm for the identification of posterior mode.\\

For each MCMC, iteration, the algorithm proceeds by calculating the misclassification of $\hat{z}$ relative to $\hat{z}^R$, and permuting the component labels of $z$ to maximise the match with $\hat{z}^R$ by calculating a misclassification cost matrix $C$, defined as
$$
C_{hj} = \{ \hat{z}_i^R=h \wedge \hat{z}_i\neq j\}, \quad i  \in 1\ldots n, \quad j,h=1,\ldots,k.
$$
Permutation of the misclassification matrix can be performed efficiently with the so-called Hungarian Algorithm (\shortciteNP{munkres57}), and the column permutation that minimises the $tr(C)$ is then recorded for each iteration of the MCMC sample.\\

\subsubsection{Papastamoulis and Iliopoulos (2010) }
\shortciteN{papastami10} introduces a similar algorithm , their algorithm can be seen as a modification of the pivotal reordering algorithm of \shortciteN{marinmr05}. The method is justified via an equivalence class representation, by redefining the symmetric posterior distribution to a nonsymmetric one
via the introduction of an equivalence class.\\

More specifically, to determine the equivalence class, a vector $z^*$ will be selected to act as a pivot, such as the posterior mode. Then for each MCMC sample output $z$, the permutation that makes $z$ as similar as possible to $z^*$ will be selected. Hence the algorithm works very similar to
\shortciteN{marinmr05} with the difference being that the similarity measure here is based on the allocation variable defined as
$$
S(z_1, z_2) := \sum_{i=1}^n I(z_{1i} = z_{2i})
$$
for two allocation vectors $z_1, z_2$, where $I(A)$ is the indicator function of $A$. \\

\section{A variance based relabelling algorithm} \label{sec:minvar}


In this section, we propose a new algorithm motivated by the expected posterior mean squared loss function,
\begin{equation}\label{eqn:loss}
L(\phi, \hat{\phi}) = \mathbb{E}_{p(\phi | x)}[(\phi-\hat{\phi})^2]= \mbox{var}(\phi) + (\mathbb{E}(\phi) - \hat{\phi})^2
\end{equation}
where $p(\phi|x)$ is the posterior distribution, thus minimising the above loss function amounts to minimizing
\begin{equation}\label{eqn:minvar}
(\nu_k^*, \hat{\phi}^*)=\underset{\nu_k, \hat{\phi}}{\mbox{argmin }}\left[\mbox{ var}(\phi_{\nu_k}) +  (\mathbb{E}(\phi_{\nu_k}) - \hat{\phi})^2\right]
\end{equation}
since for a given permutation $\nu_k$, setting $\hat{\phi}^*$ to the posterior mean minimises the second term in the above loss function.
Hence to minimise Equation (\ref{eqn:loss}),
we should find the permutation that minimises the posterior variance of the parameters.\\

In practice, exhaustive minimisation of Equation \eqref{eqn:minvar} is computationally prohibitive for large numbers of sample output. So similarly to \shortciteN{celeux98},  \shortciteN{marinmr05}, \shortciteN{cronw11} etc, we first find reference points in the modal locations, and iteratively minimize the variance of the
posterior samples with respect to the permutations in the modal region. The following proposition shows that provided that the cluster means do not change very quickly, minimisation of Equation \eqref{eqn:minvar} can be performed iteratively.\\

\begin{prop}\label{prop1}
Let $V^*_m = \sum_{i=1}^q\widehat{\mbox{var}}(\phi^{[m]}_{\nu^*, i}) $ denote the minimum total variance of the parameters $\phi^{[m]}_{\nu^*, i}$ with
corresponding optimal permutations $\nu^*$, based on $m$ iterates of the MCMC output. Let  $V^*_{m+1} =  \sum_{i=1}^q \widehat{\mbox{var}}(\{\phi^{[m]}_{\nu^{**}, i}, \phi^{(m+1)}_{\nu^{m+1},i}\}) $ denote the minimum total variance based on the sample with one additional MCMC sample, with the optimal permutations given by
$\nu^{**}$ and $\nu^{m+1}$.  Denote the parameter means by $\bar{\phi}^{[m]}_{\nu^*,i}$ and $\bar{\phi}^{[m+1]}_{\nu^{**}, \nu^{m+1}, i}, i=1,\ldots, q$. 
Suppose that $\bar{\phi}^{[m]}_{\nu^*,i} \approx \bar{\phi}^{[m]}_{\nu^{**}, i}$, then the optimal permutations $\nu^{**}= \nu^{*}$, and $V^*_{m+1}$ can be minimised by permutation of the vector $\phi^{(m+1)}$ only.
\end{prop}

\noindent{\bf Proof:} See Appendix.\\

Thus as long as the successive parameter means do not change much under optimal reordering,  
we can minimize the variance criterion iteratively, only reordering each new sample, while keeping the ordering of the previous samples unchanged. This condition
is reasonable particularly as $m$ increases.

\subsection{Minimum variance algorithm}
Here we give an algorithm based on minimising the variance of the parameters, the algorithm is based on the full parameter space, similar to those
in Section \ref{sec:fullpar}.
\begin{description}
\item {\bf \it Step 1:} Select $m$ posterior samples from the modal region, such that no switching has occurred.


\item {\bf \it Step 2:} Excluding the samples used in Step 1. For $r=1, \ldots M$,  each successive iteration of the MCMC output is relabelled
according to
$$\nu_k^{(m+r),*} = \underset{\nu_k^{(m+r)}}{\mbox{argmin}} \sum_{i=1}^q \widehat{\mbox{var}}(\{\phi^{[m+r-1]}_{\nu^*_k, i}, \phi^r_{\nu_k^{m+r},i}\}),$$
\end{description}
where $\widehat{\mbox{var}}(\phi^{[m+r-1]}_{\nu_k, i})$ is the sample variance for the $i$th parameter, under the permutation $\nu^*_k$,  corresponding to the set of
previous $m+r-1$ samples.  Relabel the $(m+r)$th sample according to $\nu^{(m+r),*}_k$.\\

In Step 1, we choose a small set of modal posterior samples, where no switching has occurred, but a good estimate of the posterior means can be obtained. This is
similar to the approach suggested in \shortciteN{celeux98}. A number between 50 to 100 is typically sufficient.
Step 2 involves only permuting the labelling of the $r$th sample to minimise the overall posterior variance including the new sample $\phi^r$. A computationally efficient update of the variance for each of the $i$th components is given by iteratively computing
$$ \bar{\phi}^{[m+r]}_i = \frac{1}{m+r}[(m+r-1)\bar{\phi}_i^{[m+r-1]} + \phi^r_i] $$
$$ \widehat{\mbox{var}} (\phi^{[m+r]}_i) =\frac{m+r-2}{m+r-1}\widehat{\mbox{var}} (\phi^{[m+r-1]}_i)  + \frac{1}{m+r}(\phi^r_i -  \bar{\phi}^{[m+r-1]}_i )^2$$
where $\bar{\phi}^{[m]}_i$ denotes the sample mean of the $i$th parameter based on $m$ samples. \\

%
%

\subsection{Simultaneous monitoring of MCMC convergence}
We note an interesting connection of the variance based relabelling algorithm with the well known Gelman and Rubin convergence assessment.  Given $J$ parallel MCMC sequences,  each with length $M$,  \shortciteN{gelman1992inference} suggest to monitor the so called potential scale reduction factor $R$ at MCMC iteration $m$, estimated as
$$\hat{R} = \sqrt{\frac{\widehat{\mbox{var}}(\phi_i )}{W}}$$
where
$$\widehat{\mbox{var}}(\phi_i ) = \frac{m-1}{m}W + \frac{1}{m}B$$
where $W$ is the within chain variance of the $i$th marginal parameter,  based on $m$ samples,
$$ W= \frac{1}{J} \sum_{j=1}^J \widehat{\mbox{var}} (\phi^{[m]}_i)$$
note that $W$ is readily given by Step 2 of the algorithm above.\\

$B$ is the between chain variance
$$B=\frac{M}{j-1} \sum_{j=1}^J (\bar{\phi}^{[m]}_{i, j} - \bar{\phi}^{[m]})^2
$$
where $\bar{\phi}^{[m]}_{i, j} $ is the sample mean of the $j$ chain, for the $i$th parameter, based on $m$ samples, and $ \bar{\phi}^{[m]} = \frac{1}{J}\sum_{j=1}^J \bar{\phi}^{[m]}_{i, j}$. Again $\bar{\phi}^{[m]}_{i, j} $  is given in Step 2 of the algorithm for a given $j$th chain. Thus the potential scale reduction factor is readily calculated, a value approaching 1 is indicative of MCMC convergence.\\

Thus to monitor the convergence of multiple MCMC sequence for each marginal parameter $i$, the above algorithm only has to be modified slightly. In Step 1, instead of selecting samples $m$ from a single chain, we will select $J$ equal sized samples $m_j, \sum_{j=1}^J m_j=m$ amongst the modal regions of the $J$ parallel chains. Then in Step 2, for each chain $j=1,\ldots J$, and their respective initial samples $m_j$, carry out Step 2 and calculate $\hat{R}$.
\\


 \section{Examples} \label{sec:example}
 In this section, we will compare all the algorithms presented above in several examples involving both real and simulated data.

\subsection{Univariate mixtures}
We first consider two univariate mixture models, a three-component and a five-component model,
\begin{equation} \label{mixture:papa1}
 0.10N(-20,1)+0.65N(20,3)+0.25N(21,0.5)
\end{equation}
\begin{equation} \label{mixture:papa2}
 0.20N(19,5)+0.20N(19,1)+0.25N(23,1)+0.20N(29,0.5)+0.15N(33,3).
\end{equation}
In the three component model, the final two components are very close together, and we expect that it will be easy to identify the first component, but not the
last two. Similarly with the five component model, the first two components will be extremely difficult to separate. This example was also studied in detail by \shortciteN{papastami10}.
\\

\begin{figure} [htp]
\centering
\subfigure[Mixture of Equation \eqref{mixture:papa1}]{\label{papa1.1}\includegraphics[width=0.95\linewidth,height=0.3\textheight]{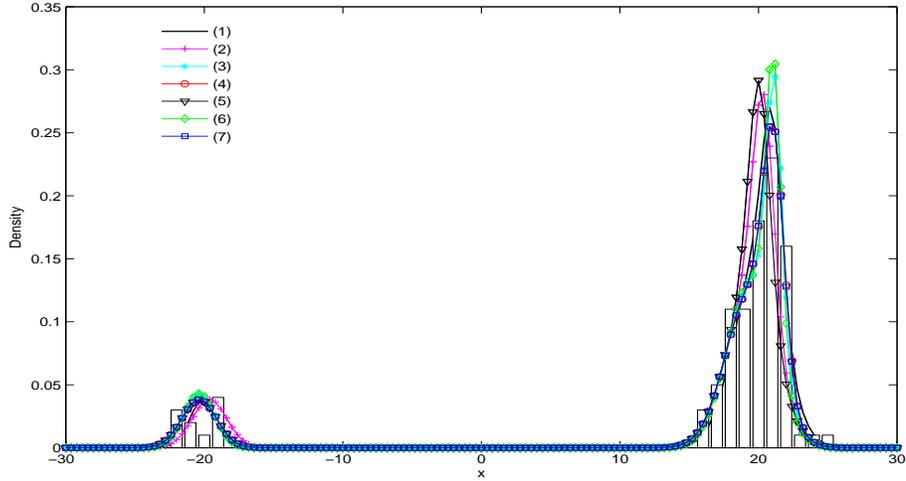}}\\
\subfigure[Mixture of Equation \eqref{mixture:papa2}]{\label{papa1.2}\includegraphics[width=0.95\linewidth,height=0.3\textheight]{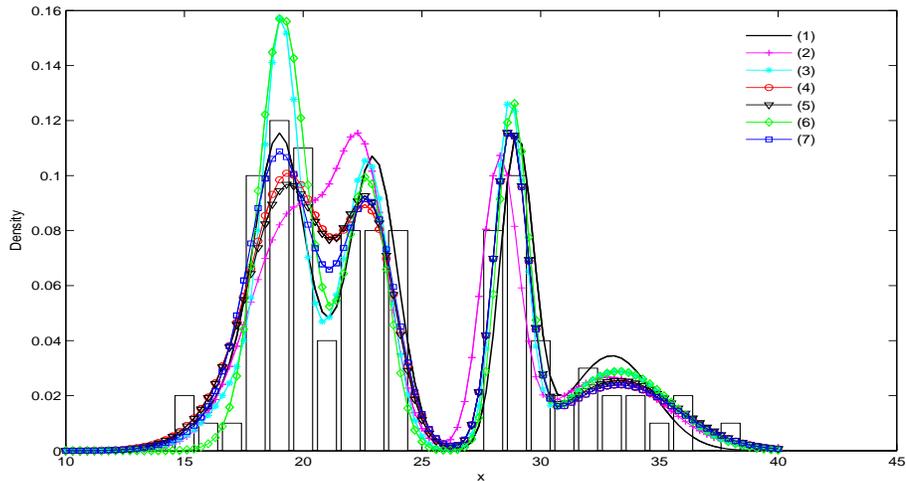}}
\caption{Histogram of 100 simulated observations from model (\ref{mixture:papa1}) and (\ref{mixture:papa2}). The superimposed lines correspond to  (1)  true density, (2) Celeux et al (3) Fr\" uwirth-Schnatter (4) Marin et al   (5) Cron and West  (6) Papastamoulis et al (7) Minimum Variance.}
\label{fig:mixture}
\end{figure}

We use a sample of 100 data points simulated from each of the two models, and follow the MCMC sampler of \shortciteN{richardson1997bayesian}. For both models,
we ran 80,000 iterations of MCMC sampling and discard the first 20,000 as burn in. Figure \ref{fig:mixture} shows the density estimates using each of the six different methods we discussed, superimposed with their true density. Clearly, all methods agree in regions where identifiability is easily separable, and differences between the different methods are more pronounced where components are very close together, and this is the case for the last two components in Model \eqref{mixture:papa1} and the first two components in Model \eqref{mixture:papa2}.
\\

Overall, with the exception of the method of Celeux et al (1998, 2000), the other methods give similar density estimates. It can be seen that in both examples, the $k$-means method of Fr\" uwirth-Schnatter (2011) is most similar to the equivalence class method of Papastamoulis et al (2010), although one is based on the full parameter space and the other is based on only the allocation variables. It can be seen that the left hand tail of Model \eqref{mixture:papa2} is under-estimated by the method of Papastamoulis et al (2010) relative to the other methods. We will return to this issue later.
\\


We find that the method of Celeux et al (1998, 2000) does not perform well in both cases. This is due to the use of a scaled distance, the process can be dominated by those components with very small variances. From the misclassification table given in Table \ref{misclassification}, we can see that the second component of Model \eqref{mixture:papa2} has been completely misclassified by the method of Celeux et al (1998, 2000), the second component was dominated by the first component.\\

Finally, we present a more thorough comparison of the six different method in Table \ref{comparison:papa}, where we give an estimate of the KL distance between the true density and the estimated densities, the overall misclassification rates (as computed in Table \ref{misclassification}), the total variance of the parameter estimates and the CPU time, where the computation was carried out using the Matlab programming language, on Ubuntu (x86\_64) with kernel version of 3.2.0-53-generic. Both the method of Cron and West (2011) and Papastamoulis et al (2010) were computed using the author's own softwares.
\\

Overall, Celeux et al (1998, 2000) has the largest KL distance, overall misclassification rate and total variance, although its computational time
is competitive with the other algorithms. We note that the method of Fr\" uwirth-Schnatter (2011) requires far more MCMC sample output than the other methods, since
samples which has been clustered into less than $K$ components has been discarded by the algorithm, hence to obtain 60,000 samples, we ran approximately 27,000 additional MCMC iterations for Equation \eqref{mixture:papa1} and an additional 90,000 iterations for  \eqref{mixture:papa2}, thus even though it is a fast algorithm
in itself, the computational overheads in the additional MCMC sampling makes this algorithm by far the most computationally
costly. In addition, although the method achieves good misclassification rate, it would appear we cannot trust the resulting parameter estimates, see Table \ref{comparison:papafull}, we believe this may be attributed to the non-random exclusion of samples from the MCMC output.
\\

The remaining methods of Marin et al (2005), Cron and West (2011), Papastamoulis et al (2010) and the proposed minimum variance algorithm, all performed relatively well. Minimum variance gave the smallest KL distance, with similar results using Marin et al (2005). The best method in terms of misclassification rate is Papastamoulis et al (2010), with Cron and West (2011) marginally worse. In terms of posterior variance,  the minimum variance algorithm produced the smallest values, closely followed by Papastamoulis et al (2010). Finally, in terms of CPU time, all methods are efficient, the best ones being Papastamoulis et al (2010) and Marin et al (2005), and the minimum variance algorithm is the slowest here.
\\

Finally, the parameter estimates given in Table \ref{comparison:papafull} show that while the mean parameters are fairly well estimated by most methods, the variance
estimates are quite different. It is clear that while Marin et al (2005), Cron and West (2011) and minimum variance all over estimated
the 2nd and the last variance components of Equation \eqref{mixture:papa2}, Papastamoulis et al (2010) underestimated the variance 
of component one while overestimating the variance of the last component. Overall, the variance estimates are generally smaller from 
Papastamoulis et al (2010) than the other three methods, and is generally underestimated relative to the true values, while the variance
estimates are generally overestimated from Marin et al (2005), Cron and West (2011) and minimum variance relative to the true values.
\\

\begin{table} [!ht]  \small 
  \centering
  \begin{tabular}{|l|ccc|ccccc|}
\hline
\hline
 &\multicolumn{3}{c}{ Eq. (\ref{mixture:papa1})} &\multicolumn{5}{c|}{ Eq. (\ref{mixture:papa2})}  \\

\hline
\multirow{5}{*}{True }         &10   & 0   & 0  & 20 & 0 & 0 & 0 & 0\\
			      &0    & 65  & 0  &0 &20  & 0 & 0 & 0\\
                               &0     & 0 & 25  &0  &0 & 25 & 0 & 0 \\
                               &      &    &     &0 & 0 & 0 & 20 & 0 \\
                               &      &    &     &0 & 0 & 0 &0 &15\\
\hline
\multirow{5}{*}{Celeux et al } &10   &0    & 0  & 18& 0 &2 &   0    &0 \\
			      &0    &43   &22  &20 & 0 &0 &   0    &0\\
                               &0    &11   & 14 &2  & 1 &22&   0    &0 \\
                               &      &    &      &0  & 0 & 0&   20   &0 \\
                               &      &    &      &0  & 0 & 0&   1    &14 \\
\hline
\multirow{5}{*}{Fr\" uwirth-Schnatter} & 10   &0    & 0    & 3 & 15 & 2 & 0 & 0\\
              & 0    &55   & 10   & 0  &  20& 0 & 0 & 0\\
                                       & 0    & 14  & 11   &0   & 2 & 23&0  &0\\
                                       &      &          &  &0   &0  & 0 &20 &0\\
                                       &      &          &  &0   & 0 & 0 & 2 &13\\

\hline
\multirow{5}{*}{Marin et al}    & 10 &0    & 0  & 6   & 12 & 2&0 &0\\
        & 0  &41   & 24 & 3   & 17 & 0&0 &0\\
                                & 0  &6    & 19 & 0   & 3  &22& 0&0\\
                                &    &     &     & 0   &  0 & 0&20&0\\
                                &    &     &      & 0   &  0 & 0&2 &13\\

\hline
\multirow{5}{*}{Cron and West} & 10   &0    & 0   &6  & 12 & 2 &   0&0\\
       & 0    &44   & 21  & 3 & 17 &  0&   0&0\\
                               & 0    &16   & 9   &0  & 3  & 22&0   &0\\
                               &      &     &       & 0 &  0 & 0 & 20 &0\\
                               &      &     &       &0  & 0  & 0 & 2  &13\\
\hline
\multirow{5}{*}{Papastamoulis et al}  & 10   &0    & 0    & 6 & 12& 2&0 &0\\
             & 0    & 42  & 23   & 0 & 20& 0&0 &0\\
                                      & 0    &6    & 19   &2  & 1 &22&0 &0\\
                                      &      &     &       &0  & 0 & 0&20&0\\
                                      &      &     &       &0  &  0& 0&2 &13\\
\hline

\multirow{5}{*}{Minimum Variance}   & 10   &0    & 0   & 6 & 12 & 2&0 &0\\
                                    & 0    & 40  & 25  &0  & 20 &0 &0 &0\\
                                    & 0    & 6   & 19  &2  & 1  &22&0 &0\\
                                    &      &     &      &0  &  0 &0 &20&0\\
                                    &      &     &       & 0 & 0  &0 &2 &13\\
\hline \hline
 \end{tabular}
 \caption{Misclassification matrix for the six methods. Each $i, j$th entry of the misclassification matrix denotes the number of observations which is classified as component $j$, while actually it belongs to component $i$. The row corresponding to True gives the true cluster membership of the observed data.}
 \label{misclassification}
 \end{table}

 \begin{table} [!htp] \small
  \centering

  \begin{tabular}{|l|cc|cc|cc|cc|}
  \hline
  \hline
&\multicolumn{2}{c|}{KL Distance} &\multicolumn{2}{c|}{Misclassification} &\multicolumn{2}{c|}{Total Variance}  &\multicolumn{2}{c|}{Time (sec)} \\
  &Eq. \eqref{mixture:papa1} &Eq. \eqref{mixture:papa2}   &Eq. \eqref{mixture:papa1} &Eq. \eqref{mixture:papa2}&Eq. \eqref{mixture:papa1} &Eq. \eqref{mixture:papa2}&Eq. \eqref{mixture:papa1} &Eq. \eqref{mixture:papa2} \\ 	
\hline
Celeux et al  &0.21 &0.38 &33\% &26\%  &62.61&82.89 &19.55 &42.69\\

Fr\" uwirth-Schnatter &0.08 &0.11 &24\% & 21\%  &1.68 & 8.89 &14.38 & 28.35 \\

Marin et al  &0.07 &0.14  & 30\% &22\% &2.30 &55.79 &17.66 &40.19 \\

Cron and West   &0.31 &0.14 &37\% & 22\% &2.38 &56.11 &26.63 &38.84\\

Papastamoulis et al &0.11 &0.35 & 29\% & 19\% &2.35 &52.96 &16.11 & 25.05\\

Minimum Variance &0.07 &0.11 & 31\%& 19\% &2.30 & 52.02&24.83 &50.98\\
\hline
\hline
 \end{tabular}
 \caption{Comparison of KL distance relative to the true distribution, misclassification rate, total variance for the parameter estimates and computation time, for the six different methods outlined, using simulated data from Equations \eqref{mixture:papa1} and \eqref{mixture:papa2} . }
 \label{comparison:papa}
 \end{table}

 \begin{table} [htp]
  \centering
  \begin{tabular}{|l|cc|cc|cc|}
\hline
\hline
 &\multicolumn{2}{c|}{ $\hat{w_k}$} &\multicolumn{2}{c|}{ $\hat{\mu_k}$} &\multicolumn{2}{c|}{ $\hat{\sigma}^2_k$ } 
 \\ 
  &Eq. \eqref{mixture:papa1} &Eq. \eqref{mixture:papa2}   &Eq. \eqref{mixture:papa1} &Eq. \eqref{mixture:papa2}&Eq. \eqref{mixture:papa1} &Eq. \eqref{mixture:papa2}\\
\hline
\multirow{5}{*}{True}    &0.10   &0.20   & -20.00& 19.00& 1.00 &5.00 \\
			&0.65   &0.20   &20.00  &19.00 & 3.00 &1.00\\
                         &0.25   &0.25   & 21.00 &23.00 & 0.50 &1.00\\
                         &       &0.20   &       &29.00 &      & 0.50\\
                         &       &0.15   &       &33.00 &      & 3.00\\

 \hline
\multirow{5}{*}{Celeux et al }& 0.12   & 0.24  &-19.63  &19.40  & 1.55      &3.45    \\
                               & 0.55   &0.19   &19.41   &20.20  & 3.13      & 4.05 \\
                               & 0.33   &0.21   &20.37   &22.58  & 0.57      &1.06   \\
                               &        &0.20   &        &28.28  &           & 0.60   \\
                               &        &0.16   &        &32.85  &           & 5.69   \\
\hline
\multirow{5}{*}{Fr\" uwirth-Schnatter} & 0.12    & 0.05    & -20.37 & 15.80   & 1.24  &0.87   \\
                                       & 0.58    &0.34      & 19.52  & 19.15   & 0.97  & 0.46    \\
                                       & 0.30    & 0.25     & 21.16  &22.83    & 0.32  & 0.88   \\
                                       &         & 0.21     &        &28.75    &       & 0.55    \\
                                       &         &0.15      &        &33.41    &       & 3.66    \\

\hline
\multirow{5}{*}{Marin et al}    & 0.12     &0.23    & -20.37 & 19.01     & 1.57   & 4.90\\
                                & 0.55     &0.21     & 19.42  & 19.43     & 3.14   & 1.98\\
                                & 0.33     &0.21     & 21.11  & 22.85     & 0.55   &1.23\\
                                &          &0.20     &        & 28.72     &        & 0.50\\
                                &          &0.15     &        & 33.30     &        & 6.25\\

\hline
\multirow{5}{*}{Cron and West} & 0.12   &0.22  & -20.37 &18.99     & 1.57      & 4.82  \\
                               & 0.56    &0.21  & 19.45  & 19.47    & 3.14      &  2.14   \\
                               & 0.32    &0.21  & 21.07  &22.84     & 0.55      & 1.14    \\
                               &         &0.20  &        & 28.72    &           & 0.50    \\
                               &         &0.16  &        &33.30     &           & 6.25    \\
\hline
\multirow{5}{*}{Papastamoulis et al}  & 0.12    & 0.18    & -20.37  & 19.51   & 1.25  & 1.93\\
                                      & 0.56    & 0.25     & 19.44   & 19.05   & 2.87  & 0.84\\
                                      & 0.32    &0.21      & 21.08   &22.73    & 0.30  & 0.76\\
                                      &         &0.20      &         &28.72    &       & 0.42\\
                                      &         & 0.16     &         &33.30    &       & 4.89\\
\hline
\multirow{5}{*}{Minimum Variance}  & 0.12     & 0.22   & -20.37 & 19.50  & 1.57   & 5.43  \\
                                    & 0.55     & 0.22   & 19.42  &18.94   & 3.14   & 1.49   \\
                                    & 0.33     & 0.21   & 21.10  &22.85   & 0.55   &1.18    \\
                                    &          & 0.20   &        &28.72   &        & 0.50 \\
                                    &          & 0.15   &        & 33.30  &        & 6.25   \\
\hline
\hline
 \end{tabular}
 \caption{Parameter estimates using the six different methods. The left part of each column corresponds to Equation \eqref{mixture:papa1}, the right part of each column corresponds to Equation \eqref{mixture:papa2}.}
 \label{comparison:papafull}
 \end{table}

\subsection{Galaxy data}
Here we compare the various methods on the well known galaxy data, which has been studied extensively in the relabelling literature,  see for example \shortciteN{stephens1997}, \shortciteN{celeuxhr00}, \shortciteN{jasraHS05}. This data set consists of the velocities of several galaxies diverging from our own galaxy. The original data set consists of 83 observations, but one of them is recorded as infinite, and so we leave this one out and use the remaining 82 observations. We follow the setup of  \shortciteN{richardson1997bayesian} in setting up the model and MCMC sampling, and fix the number of mixture components at 6, which was shown to have the highest posterior model probability. We run 80,000 MCMC iterations and discard the first 20,000 iterations, keeping the final 60,000 samples. For the method of
Fr\" uwirth-Schnatter (2011), we ran an additional 320,000 iterations.\\

\begin{figure}[htp]
  \centering
\includegraphics[width=0.8 \textwidth,height=0.3 \textheight]{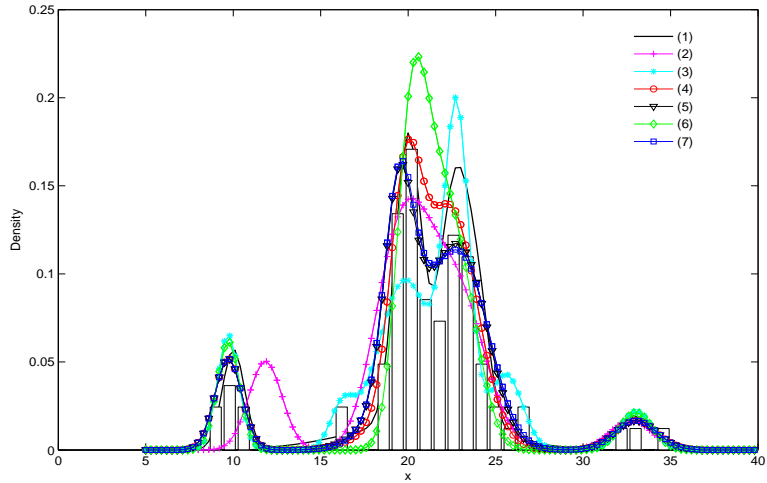}%
  \caption{Histogram of the galaxy data.  The superimposed lines correspond to  (1)  posterior MAP density estimate (2) Celeux et al (3) Fr\" uwirth-Schnatter (4) Marin et al   (5) Cron and West  (6) Papastamoulis et al (7) Minimum Variance.}
  \label{galaxy:hist}
 \end{figure}

 \begin{table}
  \centering
  \begin{tabular}{|l|c|c|c|}
  \hline
  \hline
&KL Distance  &Total Variance  &Time (sec) \\
\hline
Celeux et al  &2.94 &91.87 &2.94\\

Fr\" uwirth-Schnatter  &1.89 &8.89 &71.70  \\

Marin et al  &1.37 &37.57 &40.45 \\

Cron and West  &1.40&40.07&49.79  \\

Papastamoulis et al &2.61&52.06&29.35 \\

Minimum Variance &1.30&38.52   &87.70 \\
\hline
\hline
 \end{tabular}
 \caption{Comparison of KL distance relative to the MAP density estimate,  total variance for the parameter estimates and computation time, using the six different methods,  for the galaxy data.  }
 \label{comparison:galaxy}
 \end{table}

Figure \ref{galaxy:hist} shows histogram and density estimate of galaxy data.  Here the differences between the methods are more pronounced than in the
previous example. Again, it is clear that both Celeux et al (1998, 2000) and Fr\" uwirth-Schnatter (2011) are not performing well. Both the figure and Table \ref{comparison:galaxy} show that the method of Marin et al (2005), Cron and West (2011) and minimum variance are the most similar to
each other, and give smaller total variance estimates.  Papastamoulis et al (2010)  was the most efficient in terms of computing time, suggesting that
the method scales up well with the number of components. Finally, again we see from the table of parameter estimates in Table \ref{galaxy:result} that the 
estimates of variances are generally smaller for Papastamoulis et al (2010)  than the other methods.
\\

  \begin{table} [ht] \footnotesize 
  \centering
  \begin{tabular}{|l|c|cccccc|}
\hline
 \hline
 \multirow{3}{*}{MAP}     &{$\hat{w}_k$}           &0.09  &0.31   &0.15   &0.33   &0.07   &0.05 \\
    		&{$\hat{\mu}_k$}  &10.01 &20.00  &20.60  &22.76  &24.14  &32.86\\
                      & $\hat{\sigma}^{2}_k$ &0.40  &0.57  &16.38  &0.84   &0.42  &1.08  \\
                       \hline

\multirow{3}{*}{Celeux et al} &$\hat{w}_k$           &0.12  & 0.15  & 0.27  & 0.18  & 0.23  & 0.05   \\
		&$\hat{\mu}_k$        &11.84 & 18.73 & 20.19 & 21.74 & 22.95 & 32.72\\
                                              & $\hat{\sigma}^{2}_k$ &0.90  & 1.48  & 1.75  & 4.19  & 2.25 & 1.41  \\

\hline
\multirow{3}{*}{ Fr\" uwirth-Schnatter}   &$\hat{w}_k$            &0.10   & 0.04   & 0.35   & 0.35   & 0.10   & 0.05   \\
				&$\hat{\mu}_k$        &9.71 & 16.38    & 19.79  & 22.58  & 25.48  & 33.03  \\
                                            &$\hat{\sigma}^{2}_k$  &0.37   & 0.50  & 0.47     & 1.14   & 0.76   & 0.81  \\

 \hline
\multirow{3}{*}{Marin et al}    &$\hat{w}_k$           &0.10 & 0.22 & 0.20 & 0.22 & 0.20 & 0.05 \\
	 &$\hat{\mu}_k$            &9.72 & 19.46 & 20.40 & 22.12 & 23.46 & 33.02\\
                                          & $\hat{\sigma}^{2}_k$ &0.59 & 0.65 & 4.20 & 3.30 & 1.81 & 1.42\\
\hline
\multirow{3}{*}{Cron and West}  &$\hat{w}_k$           &0.10 & 0.27 & 0.16 & 0.16 & 0.27 & 0.05 \\
			 &$\hat{\mu}_k$        &9.72 & 19.86 & 20.71 & 22.15 & 22.72 & 33.02\\
                                            & $\hat{\sigma}^{2}_k$ &0.59 & 0.76 & 4.94 & 2.26 & 2.00 & 1.42 \\

\hline
\multirow{3}{*}{Papastamoulis et al}    &$\hat{w}_k$          &0.10 &0.25 &0.20 &0.13 &   0.27  & 0.05\\
		&$\hat{\mu}_k$        &9.72 &20.00&20.79&21.74&    22.90& 33.02\\
                                           &$\hat{\sigma}^{2}_k$&0.42 &0.63 &0.76 &0.92 &    1.33 & 0.89 \\

\hline
\multirow{3}{*}{Minimum Variance}       &$\hat{w}_k$        &0.10 & 0.22    & 0.20   & 0.22 & 0.21 & 0.05 \\
		&$\hat{\mu}_k$           &9.72 & 19.46   & 20.38 & 22.08 & 23.52 & 33.02\\
                                        & $\hat{\sigma}^{2}_k$ &0.59 & 0.68    & 3.63    & 3.67 & 1.98 & 1.42 \\
\hline
\hline
 \end{tabular}
 \caption{Parameter estimates for galaxy data using different relabelling algorithms and the MAP estimate.}
 \label{galaxy:result}
 \end{table}

\subsection{High dimensional image segmentation example}

We consider a multivariate spatial mixture model in the context of image analysis, where the both the dimension of the mixture, as well as the dataset itself can be large.
Here we use a simulated  3-D image of $50 \times 50\times 16$  voxels, this is equivalent to having 40,000 observations. 
We assume that  each voxel comes from a 3 dimensional mixture model of two components, the mean parameters are $\mu_{1}=[4,5,6]$ and $\mu_{2}=[6,7,8]$ respectively. The corresponding  covariance matrices are:
\begin{equation*}
  0.5\times\left( \begin{array}{ccc}
1.00 & 0.80 & 0.64 \\
0.80 & 1.00 & 0.80 \\
0.64 & 0.80 & 1.00 \end{array} \right),  \textrm{and} \
   0.5\times\left( \begin{array}{ccc}
1.00 & 0.50&0.25 \\
0.50 & 1.00 & 0.50\\
0.25 & 0.50 & 1.00 \end{array} \right).
\end{equation*}
In real applications, such as in  dynamic positron emission tomography (PET), or  functional MRI studies, the number of observations and the dimensions of the mixture is much larger.
This example demonstrates the need for fast and reliable relabelling algorithms.\\

\begin{figure}[htp]
  \centering
\begin{tabular}{cc}
{\includegraphics[width=0.45\linewidth,height=0.3\textheight]{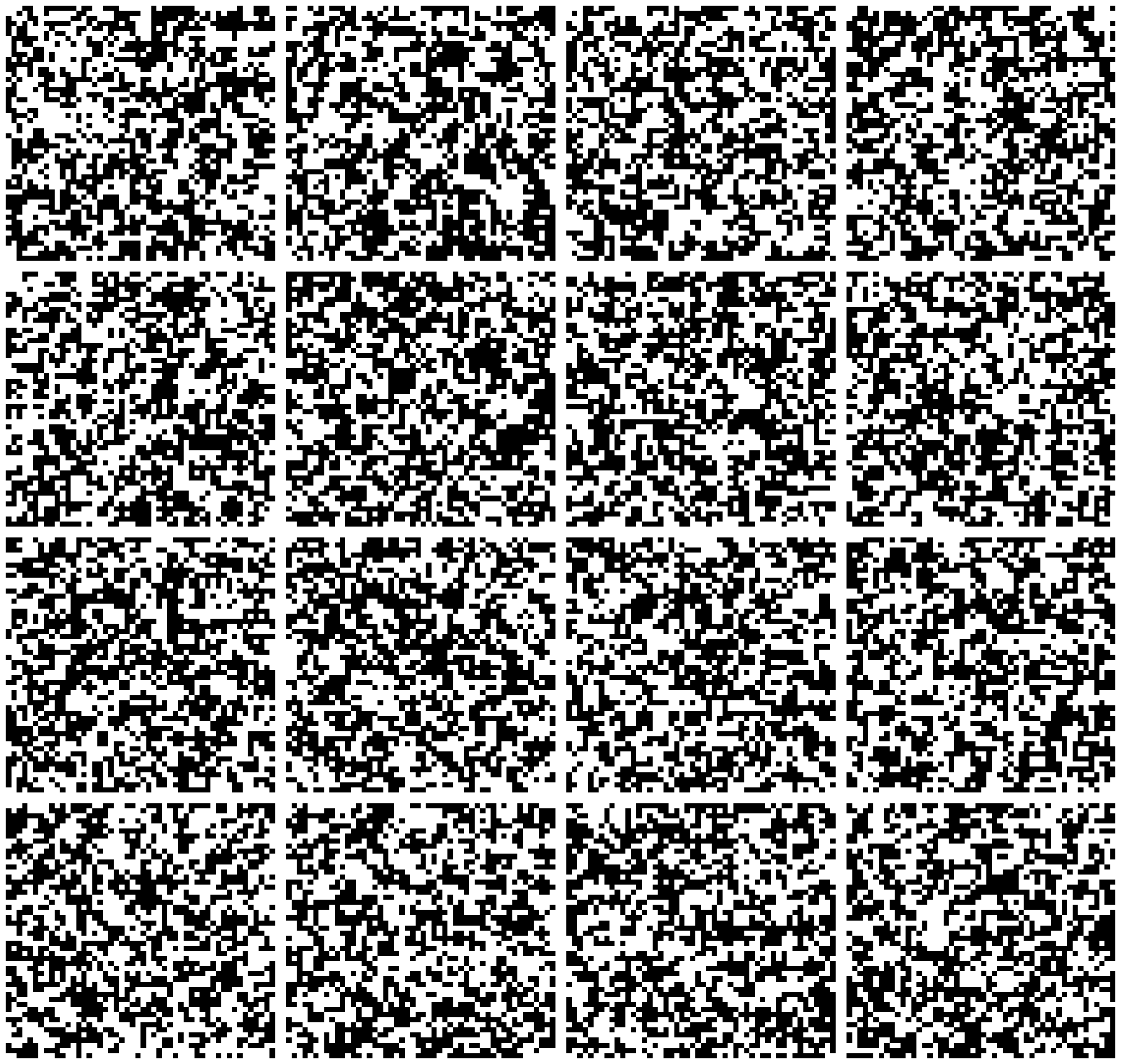}}
&{\includegraphics[width=0.55\linewidth,height=0.3\textheight]{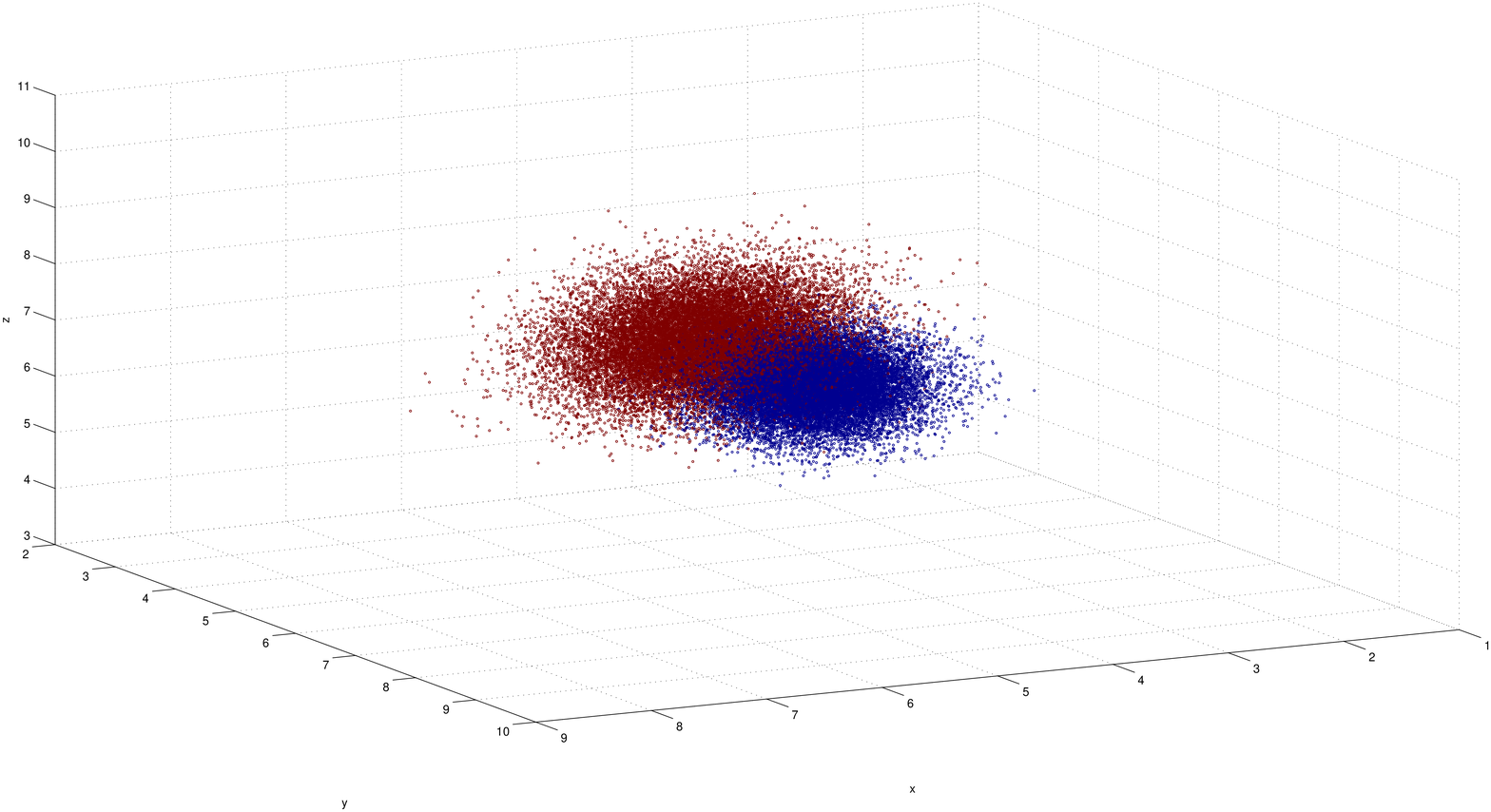}}
\end{tabular}
 \caption{The true allocations shown slice by slice (left). The white points correspond to the component with $\mu=[4,5,6]$; and black ones denote the component with $\mu=[6,7,8]$ and, 3D scatter plot of the two components (right).}
  \label{allocation}
 \end{figure}

To simulate a spatially dependent image, we first simulate the voxels using a Potts (or Ising in the case of two component mixtures) model, with spatial correlation parameter set to 0.3 ( \shortciteNP{feng2012mri}), and then assign voxel values according to the component Normal distributions. See Figure \ref{allocation} for a plot of the true allocations.\\

In the spatial clustering model, we have
\begin{equation}\label{equSpatialAlloc}
 p(x_i|  {\bphi}, z_i)=\sum_{k=1}^{K}f(x_i \mid {\bf \theta}_{k}, z_i)p(z_i=k)
\end{equation}
where the distribution of the allocation variables is given by the Potts model, 
\begin{equation}\label{equ 1.1}
  P(z | \kappa)=\frac{1}{c(\kappa)}\exp\{\kappa \sum_{i \in \delta_j } I(z_i=z_j)\}
\end{equation}
where $\delta_j$ denotes the neighbourhood of $j$, and $\kappa$ denotes the strength of the spatial connectedness (c.f. Equation \eqref{equ1.2}). 
The normalising constant $c(\kappa)$ is intractable, and we follow  \shortciteN{fernandezg02} and \shortciteN{smith2006estimation} in precomputing these in a look up table.
\\

We set the prior for $\kappa$ to be a truncated Normal(0.6,100) on the interval $[0,1] $, and use conjugate priors for the mean parameter $\mu_k  | \Sigma_k \sim N(0,  100\times\Sigma_k)$ and
covariance matrices follow an Inverse-Wishart distribution $\Sigma_k \sim IW(3, 1.5\times I_{3\times3})$, for $k=1,\ldots,K$. A hybrid Gibbs within Metropolis  sampler can be constructed from the full conditional distributions, and convergence of the MCMC sampler is obtained after 10,000 iterations, discarding the initial
5,000 samples as burn in.
In order to guarantee the presence of the label switching phenomenon, we manually switch the samples during simulation, see  \shortciteN{papaspiliopoulos2008retrospective} and \shortciteN{jasraHS05}.
\\

Table \ref{multivariate mixture} provides the posterior mean estimates for the model parameters using different reordering schemes. Here all the methods worked well, as the two mixture components are fairly well separated in the example. Table \ref{comparison:multi} gives the comparative KL divergence, misclassification rates, total variance and computing time.  Here, due to the large size of the data, small differences in the posterior parameter estimates for the method of Celeux et al translated into a relatively large KL measure. The other methods are all comparable in terms of misclassification and total variance. In terms of computational time, with the
allocation based methods taking longer than the full parameter based methods.
This example illustrates that parameter based algorithms scale up better when the size of the data increases, since the corresponding increase in the number of allocation variables needed do not affect the efficiency in the relabelling algorithms. On the other hand, if the dimension of the parameter space increase, i.e, as the dimension of the multivariate Normals increase, we would expect the allocation based relabelling algorithms to be more efficient.

\begin{table}[htp] \footnotesize
  \centering
  \begin{tabular}{|l|c|cc|}
\hline
\hline
 & \multicolumn{1}{c|}{$\hat{\mu}_k$} & \multicolumn{2}{c|}{$\hat{\sigma}_k^{2}$}\\ \hline
 True&$\left( \begin{array}{ccc}
  4.00 & 5.00 & 6.00 \\
 6.00 & 7.00& 8.00 \\ 
 \end{array} \right)$&$\left( \begin{array}{ccc}
 0.50 & 0.40 & 0.32 \\
 0.40 & 0.50 & 0.40 \\
 0.32 & 0.40 & 0.50 \end{array} \right)$&$\left( \begin{array}{ccc}
 0.50 & 0.25 & 0.13 \\
 0.25 & 0.50 & 0.25 \\
 0.13 & 0.25 & 0.50 \end{array} \right)$\\
 \hline
 Celeux et al&$\left( \begin{array}{ccc}
 4.05 & 5.05 & 6.05 \\
 
 5.97 & 6.97 & 7.97 \\

 \end{array} \right)$ 
&$\left( \begin{array}{ccc}
 0.51 & 0.41 & 0.32 \\
 
 0.41 & 0.52 & 0.41 \\

 0.32 & 0.41 & 0.51 \\

 \end{array} \right)$ &$\left( \begin{array}{ccc}
 0.50 & 0.25 & 0.13 \\

 0.25 & 0.50 & 0.25 \\

 0.13 & 0.25 & 0.50 \\

 \end{array} \right)$ \\ \hline
 Fr\"{u}wirth-Schnatter& $\left( \begin{array}{ccc}
 4.01 & 5.01 & 6.01 \\

 6.01 & 7.01 & 8.01 \\

 \end{array} \right)$
 &$\left( \begin{array}{ccc}
 0.51 & 0.41 & 0.33 \\

 0.41 & 0.52 & 0.41 \\

 0.33 & 0.41 & 0.51 \\
 
 \end{array} \right)$ & $\left( \begin{array}{ccc}
 0.50 & 0.25 & 0.11 \\

 0.25 & 0.49 & 0.24 \\

 0.11& 0.24 & 0.50 \\

 \end{array} \right)$\\ \hline
 Cron and West &$\left( \begin{array}{ccc}
4.01 & 5.01 & 6.01 \\
 6.01 & 7.01 & 8.01 \\

 \end{array} \right)$ 
 
  & $\left( \begin{array}{ccc}
 0.51 & 0.41 & 0.33 \\
 
 0.41 & 0.52 & 0.41 \\

 0.33 & 0.41 & 0.51 \\

 \end{array} \right)$ &$\left( \begin{array}{ccc}
 0.50 & 0.25 & 0.12 \\

 0.25 & 0.50 & 0.24 \\

 0.12 & 0.24 & 0.51 \\

 \end{array} \right)$  \\ \hline
 Marin et al &$\left( \begin{array}{ccc}
 4.01 & 5.01 & 6.01 \\
 6.01 & 7.01 & 8.01 \\
 \end{array} \right)$ 
 & $ \left( \begin{array}{ccc}
 0.51 & 0.41 & 0.33 \\
 0.41 & 0.52 & 0.41 \\
 0.33 & 0.41 & 0.51 \\
 \end{array} \right) $  & $\left(\begin{array}{ccc}
 0.50 & 0.25 & 0.12 \\
 0.25 & 0.49 & 0.24 \\
 0.12 & 0.24 & 0.50 \\

 \end{array} \right)$\\ \hline
 Papastamoulis et al &$\left( \begin{array}{ccc}
 4.01 & 5.01 & 6.01 \\
 6.01 & 7.01 & 8.01 \\
 
 \end{array} \right)$
&$\left( \begin{array}{ccc}
 0.51 & 0.41 & 0.33 \\

 0.41 & 0.52 & 0.41 \\

 0.33& 0.41 & 0.51 \\

 \end{array} \right)$&$\left( \begin{array}{ccc}
 0.50 & 0.25 & 0.12 \\

 0.25 & 0.49 & 0.24 \\

 0.12 & 0.24 & 0.50 \\

 \end{array} \right)$\\ \hline

 Minimum Variance &$\left( \begin{array}{ccc}
4.01 & 5.01 & 6.01 \\
 6.01 & 7.01 & 8.01 \\

 \end{array} \right)$
&$\left( \begin{array}{ccc}
 0.51 & 0.41 & 0.33 \\

 0.41 & 0.52 & 0.41 \\

 0.33 & 0.41 & 0.51 \\

 \end{array} \right)$&$\left( \begin{array}{ccc}
 0.50 & 0.25 & 0.12 \\

 0.25 & 0.50 & 0.24 \\

 0.12 & 0.24 & 0.50\\

 \end{array} \right)$\\ 
 \hline
 \hline
 \end{tabular}
 
 \caption{Posterior mean estimates of the two-components multivariate spatial mixture model, for the six different methods.}
 \label{multivariate mixture}
 \end{table}

\begin{table}
  \centering
  \begin{tabular}{|l|c|c||c|c|}
  \hline
  \hline
&KL  & misclassification&Total Variance  &Time (sec) \\
\hline
Celeux et al  &4300.80& 7.51\% &0.48 &3.00\\

Fr\" uwirth-Schnatter  &38.87&7.50\% &$8.005*10^{-4}$ & 0.39  \\

Marin et al  &38.03 &7.50\%&$8.008*10^{-4}$ &0.48 \\

Cron and West  &37.80&7.50\%&$8.008*10^{-4}$&113.26  \\
Papastamoulis et al &25.93&7.50\%&$8.008*10^{-4}$&18.46\\

Minimum Variance &38.83& 7.50\%&$8.008*10^{-4}$  &1.39 \\

%
%
%
%
%
\hline
\hline
 \end{tabular}
 \caption{Comparison of KL divergence,  misclassification rates,  total variance and computing time for the six different methods. The multivariate spatial mixture model.}
 \label{comparison:multi}
 \end{table}

\section{Summary and Conclusion}\label{sec:discussion}

In this paper, we introduced a new algorithm based on a loss function argument.  We also comprehensively compared the new algorithm with some existing relabelling algorithms, restricting our comparison to those algorithms which are scalable to higher dimensions. Where applicable, we computed the KL divergence, the 
misclassification rate, the total variance of the posterior parameter estimates and computing time, based on several examples including univariate mixtures and 
multivariate spatial mixture models, as well as on a real data set.\\

We found that the method of Celeux et al (1998, 2000) can be very sensitive, and does not always perform well, and the method of \shortciteN{schnatter11} requires much more additional MCMC sampling in some cases, so we do not recommend these two methods as a generic relabelling algorithm. The performance of the remaining four methods are similar, in terms of the criterions we used. All these methods have performed well, under the different conditions, however, all the methods give slightly different solutions. \\

In terms of performance, we can broadly group the method of \shortciteN{marinmr05}  and our proposed minimum variance algorithm. Both are based on full
parameter vectors, and show comparable performance in all the simulations we have considered. The other two, the method of \shortciteN{cronw11}  and \shortciteN{papastami10} are based on allocation variables. Although all four methods produced similar results, the method of  \shortciteN{papastami10} tended to produce an underestimated variance parameter estimate, while the other three produced an overestimated variance. Broadly speaking, the full parameter methods are more efficient for large datasets and the allocation methods are more efficient when the parameter space is large. From a more theoretical perspective, while \shortciteN{marinmr05}  simply use the canonical scalar product as an optimisation criterion, the minimum variance algorithm minimises the expected posterior loss, while
\shortciteN{cronw11}  minimises the misclassification matrix and the algorithm of \shortciteN{papastami10} is justified by an equivalence class representation. Thus from a theoretical perspective, the minimum variance algorithm and \shortciteN{papastami10} is more satisfying. We summarise the above discussion in Table \ref{comparison:final}.\\

\begin{table} \footnotesize
  \centering
  \begin{tabular}{|l|c|c|c|c|c|c|}
  \hline
  \hline
  &Optimisation criterion  &Scalability in &potential issues\\
\hline
Marin et al  &scalar product&number of data points&overestimation of variance \\

Cron and West  & misclassification  &number of parameters&overestimation of variance  \\

Papastamoulis et al &equivalence class &number of parameters&underestimation of variance \\

Minimum Variance &expected squared loss&number of data  points&overestimation of variance\\
\hline
\hline
 \end{tabular}
 \caption{Summary of the main points for the four methods, Marin et al, Cron and West, Papastamoulis et al  and minimum variance.}
 \label{comparison:final}
 \end{table}

Finally, we note that in practice, all methods can fail to find the correct labelling, see \shortciteN{cronw11}. Our simulation comparisons highlight the difficulty in distinguishing a clearly superior algorithm. From a practical perspective, we found four of the algorithms (including a novel approach introduced in this article) have similar performance, and the user may base their choice on computational considerations.

\newpage
\section*{Appendix}
\noindent{\bf Proof of Proposition \ref{prop1}}\\
Let  $V^*_m(\phi_{\nu^*}^{[m]} )= \sum_{i=1}^q\widehat{\mbox{var}}(\phi^{[m]}_{\nu^*, i}) $ denote the minimum total variance of the parameters $\phi^{[m]}_{\nu^*, i}$ with
corresponding optimal permutations $\nu^*$, based on $m$ samples. Suppose we have an additional sample $m+1$, then 
\begin{eqnarray*}
V_{m+1}(\phi_{\nu}^{[m+1]}) &=&\sum_{i=1}^q \widehat{\mbox{var}}(\phi^{[m+1]}_{\nu, i}) \\
&=&\sum_{i=1}^q\left[  \frac{m-1}{m} \widehat{\mbox{var}}(\phi_{\nu,i}^{[m]}) +\frac{1}{m+1}(\phi^{(m+1)}_{\nu, i} - \bar{\phi}^{[m]}_{\nu,i})^2\right]
\end{eqnarray*}

Then the first term inside the bracket is minimised at $\nu=\nu^*$.  
In addition, 
since we assume that, 
$\bar{\phi}^{[m]}_{\nu^{**},i}\approx \bar{\phi}^{[m]}_{\nu^*,i}$, where $\nu^{**}$ denote the optimal ordering of the $m+1$ samples. 
That is, since we assume that the component means do not change much at successive iterations, 
we can minimise the second term by minimising
$(\phi^{(m+1)}_{\nu, i} - \bar{\phi}^{[m]}_{\nu^*,i})^2 $.
Consequently, to minimize $V_{m+1}(\phi_{\nu}^{[m+1]}) $,  we only need to minimize the variance with respect to the permutations of the vector $\phi^{(m+1)}$.

\newpage
\bibliographystyle{chicago}
  \bibliography{labelswitch}
\end{document}